# Two- and three-dimensional magnetic correlations in the spin-1/2 square-lattice system $Zn_2VO(PO_4)_2$

S. M. Yusuf,[1,*] A. K. Bera,[1] N. S. Kini,[2,†] I. Mirebeau,[3] and S. Petit[3]

[1]Solid State Physics Division, Bhabha Atomic Research Centre, Mumbai 400085, India

[2]Chemistry and Physics of Materials Unit, Jawaharlal Nehru Centre for Advanced Scientific Research, Jakkur, Bangalore 560 064, India

[3]Laboratoire Léon Brillouin, CEA-CNRS, CEA/Saclay, 91191 Gif sur Yvette, France

**ABSTRACT**

The magnetic correlations in the quasi-two dimensional spin-1/2 square-lattice system $Zn_2VO(PO_4)_2$ have been investigated by neutron diffraction technique. A long-range antiferromagnetic (AFM) ordering below 3.75 K ($T_N$) has been observed with a reduced moment of 0.66(2) $\mu_B$ per V ion at 1.5 K. In a given *ab* plane, the AFM spin arrangement is Néel type and the AFM layers are coupled ferromagnetically along the *c* axis. Remarkably, we have observed a pure 2D short-range AFM ordering in the *ab* plane above $T_N$. Interestingly, the coexistence of diffuse magnetic scattering and three dimensional antiferromagnetic Bragg peaks has been found below $T_N$, indicating the presence of spin-waves as confirmed by our calculation using the linear spin-wave theory. The observed results are discussed in the light of existing theory for a two-dimensional spin-1/2 square-lattice system.

## I. INTRODUCTION

Low dimensional magnetism is one of the key areas of interest in the condensed matter physics. Low (spatial) dimensional magnetic materials show novel non-classical behaviour, induced by quantum fluctuations. The reduced dimensionality of the magnetic ordering and its crossover to a higher dimension can be understood either by strongly anisotropic exchange interactions, or by the strongly frustrating topology of the systems. The presence of a long-range magnetic order at a finite temperature in low dimensional systems ($D < 3$) was observed experimentally in presence of non-isotropic exchange constants, single ion anisotropy, and/or infinite-range interactions *e.g.*, the RKKY interactions, classical dipolar interactions, *etc* .[1-3] Each of these factors breaks the Mermin-Wagner conditions[4] and may result into a long-range order at $T << JS^2$ and a short-range order above the long-range ordering temperature (retained up to $T \sim JS^2$), where $J$ is the exchange integral and $S$ is the spin.[3]

A large number of low dimensional magnetic materials have been investigated in literature. For example, the low dimensional transition metal oxides have attracted considerable attention after the discovery of high-temperature superconductivity in layered cuprates whose parent compound provides a model 2D spin-1/2 (quantum) Heisenberg antiferromagnet on a square-lattice. The vanadium phosphates of the type $AA'VO(PO_4)_2$ [where, $A, A'$ = Pb, Ba, Sr, Zn, Cd, and Ag] have been studied extensively in the recent years as model compounds for frustrated square-lattice systems.[5] These vanadium phosphates are also interesting in the field of low dimensional magnetism because of their in-plane-like (2D) magnetic ordering arising from their intrinsic layered type crystal structures.[5, 6] The quasi-2D spin-1/2 system $Zn_2VO(PO_4)_2$ is of particular interest due to its unusual physical properties.[6] Specific heat and dc susceptibility study suggested the presence of a three dimensional antiferromagnetic (3D AFM) ordering below the Néel temperature ($T_N$ = 3.75 K) and short-range spin-spin correlations above $T_N$.[6] For a microscopic understanding of the magnetic ground state as well as short-range spin-spin correlations and the crossover to 3D long-range AFM ordering, a neutron diffraction study on this quasi-2D system is necessary.

In this paper, we report for the first time a detailed neutron powder diffraction study on $Zn_2VO(PO_4)_2$. The evolution of 3D long-range and 2D short-range magnetic correlations between the vanadium spins was deduced from the temperature dependence of the magnetic Bragg peaks and diffuse scattering in the temperature range of $1.5 \leq T \leq 7$ K. The diffuse scattering which coexists with the magnetic Bragg peaks below $T_N$ was attributed to the presence of spin-waves below $T_N$. The contribution of spin-waves to the magnetic scattering was calculated using the linear spin-wave theory, and compared to the experimental data. The observed results are interpreted in the light of the theoretical predictions for such class of systems .[7-9]

## II. EXPERIMENTAL

The polycrystalline sample of $Zn_2VO(PO_4)_2$ was prepared by the solid state reaction method as described in Ref. 6. The single phase nature of the sample was confirmed by the neutron powder diffraction (NPD) study at 300 K at the Dhruva reactor, Trombay, INDIA. For the magnetic study, low temperature (1.5-15 K) neutron diffraction measurements were carried out using the G6.1 powder neutron diffractometer at LLB, Saclay, France, covering a limited range of scattering vector length of ~ 0.12–2.5 Å$^{-1}$. The Rietveld method, by using the FullProf Program[10], was employed to refine the crystal structure and determine the magnetic structure .

## III. RESULTS AND DISCUSSION

The Rietveld refined NPD patterns at 300 and 1.5 K are shown in Figs. 1(a) and 1(b), respectively. The compound crystallizes in the tetragonal symmetry with the space group *I4cm* with lattice constants *a* = 8.9286 (7) Å and *c* = 9.0392 (1) Å. The crystal structure of $Zn_2VO(PO_4)_2$ [Fig. 2(a)] consists of a stacking of $VO_5$ square pyramidal layers, separated by a layer of $ZnO_5$ square pyramids, along the crystallographic *c*-direction. Within a given layer of $VO_5$, the pyramids of $VO_5$ are arranged on a square-lattice and connected through corner sharing $PO_4$ tetrahedra [Fig. 2(b)]. In contrast, in a given layer of $ZnO_5$, the square pyramids of $ZnO_5$ share their edges making a dimmer-like structure [Fig. 2(b)]. The appearance of extra Bragg peaks at 1.5 K in Fig. 1(b) at the scattering angles (2θ) 30.7, 72.6, 103.9, and 105.1 deg. ($Q$ ~ 0.70, 1.57, 2.09 and 2.11 Å-1, respectively), where the nuclear Bragg reflections are forbidden for the space group I4cm, confirms a 3D long range (LR) AFM ordering of the V moments. The extra (magnetic) Bragg peaks were indexed as (100), (120), (212), and (300) with respect to the tetragonal chemical unit cell. The size of the magnetic unit cell is found to be same as the chemical unit cell. The temperature dependence of the V magnetic moment, plotted in the lower inset of Fig. 1(b), indicates that the 3D AFM ordering persists up to a temperature of about 3.75 K ($T_N$). The same scale factors were used for both nuclear and magnetic phases in the Rietveld refinement. The analysis of all NPD patterns, collected at temperatures 1.5, 2.0, 2.5, 3.0, and 3.5 K (*i.e.*, at $T < T_N$), reveals that in a given *ab* plane, each V moment (aligned along the *c* axis) is coupled antiferromagnetically to its four nearest neighbor V moments (Néel AFM state) [Fig. 3]. Such AFM layers are coupled ferromagnetically along the *c* axis. To ascertain this result, we have shown in Fig 1 (c), the magnetic pattern at 1.5 K obtained by subtracting the pattern at 15 K (paramagnetic state), along with the magnetic patterns calculated by assuming either ferromagnetic or AFM coupling along the *c* direction. Clearly, the pattern calculated by considering a FM inter-layer coupling fits the observed data. For an AFM inter-layer coupling, magnetic peaks are expected at $Q \sim 0.99$, 1.72, and 2.22 Å$^{-1}$ positions [Fig 1(c)] which can be indexed as (110), (211), and (310), respectively. This is in contrast with the observed magnetic Bragg peaks at $Q \sim 0.70$, 1.57, 2.09 and 2.11 Å$^{-1}$. This concludes the presence of a ferromagnetic coupling of the AFM *ab* planes along the crystallographic *c* direction. At 1.5 K, the effective site-averaged ordered magnetic moment is derived to be 0.66(2) $\mu_B$ per V ion which is almost exactly the moment expected ($\sim 0.6$ $\mu_B$ per V ion) in a pure 2D antiferromagnetic square lattice with $S = 1/2$, which is decreased by quantum fluctuations and spin waves.[11, 12] The observed slightly higher value of the ordered moment (0.66 $\mu_B$) may arise from the interlayer ferromagnetic exchange coupling which tends to stiffen the magnetic lattice, reducing the influence of the magnetic fluctuations.

The magnetic diffraction patterns at 10, 7, 4.5, and 4 K, obtained by subtracting a diffraction pattern measured at 15K in the paramagnetic state, are shown in Figs. 4(a)-(d), respectively. In the temperature range $4 \leq T$ (K) $\leq 7$, a broad asymmetric peak appears at the same $Q$ position($\sim 0.70$ Å$^{-1}$) where the most intense 3D magnetic Bragg peak (100) is observed below $T_N = 3.75$ K. The observed asymmetric powder diffraction profile is a typical signature of 2D spin-spin correlations.[13-15] For Zn$_2$VO(PO$_4$)$_2$ with layered structure, when the effective magnetic exchange coupling along the *c* axis is very weak compared to that in the *ab* plane, 2D magnetic correlations are expected above $T_N$. A sharp symmetric magnetic Bragg peak or a Lorentzian-type broad peak profile should be expected for 3D long-range magnetic ordering or 3D short-range magnetic ordering, respectively. In the present study, an asymmetric profile (saw-tooth type) of magnetic Bragg peak, with a sharp increase in intensity at the scattering angle $2\theta_B$ and a slow decrease at higher scattering angles, is observed. The scattered intensity of a 2D Bragg reflection (*hk*) can be expressed by the Warren function[16] as,

$$I_{hk}(2\theta) = C \left[ \frac{\xi_{2D}}{(\lambda\sqrt{\pi})} \right]^{1/2} j_{hk} |F_{hk}|^2 \frac{(1+\cos^2 2\theta)}{2(\sin^{1/2}\theta)} F(a) \qquad (1)$$

where, $C$ is a scale factor, $\xi_{2D}$ is the 2D spin-spin correlation length within the 2D layer, $\lambda$ is the wavelength of the incident neutrons, $j_{hk}$ is the multiplicity of the 2D reflection (*hk*) with 2D magnetic structure factor $F_{hk}$, and $2\theta$ is the scattering angle. The function $F(a)$ is given by $F(a) = \int_0^\infty \exp\left[-(x^2-a)^2\right] dx$ where, $a = (2\xi_{2D}\sqrt{\pi}/\lambda)(\sin\theta - \sin\theta_{2DB})$ and $\theta_{2DB}$ is the Bragg angle for the 2D (*hk*) reflection. In the present case, the Warren function yields a good fit of the observed intensity profile, but not the Lorentzian function. The expression (1) was calculated numerically to get the best possible agreement with the observed magnetic diffraction patterns in the temperature range of 4-7 K over the *Q* range of 0.25-1.2 Å$^{-1}$. The calculation yields $\xi_{2D}$ values of 33(2), 28(2), 25(2) and 21(2) Å at 4, 4.5, 5 and 7 K, respectively. It is, therefore, evident that the in-plane (*ab*) 2D AFM correlation length is relatively short-ranged and decreases steadily with increasing temperature. At *T* > 7 K, the Warren scattering becomes too weak to be estimated in the present neutron diffraction study.

The occurrence of the 2D short-range AFM ordering above $T_N$ can be ascribed to the intrinsic layered type crystal structure (alternative stacking of VO$_5$ and ZnO$_5$ pyramidal layers along the *c*-axis) [Fig. 2(a)]. Here, in a given *ab* plane, the shortest possible path-way for an AFM superexchange interaction is a four-bond path-way via O1 (basal oxygen of VO$_5$ square pyramid) and P ions i.e., V-O1-P-O1-V whereas, in the out-of-plane, it is a six-bond path-way through P-O2-Zn bridge i.e., V-O1-Zn-O2-P-O1-V (Fig. 5). Therefore, due to the strongly different exchange path-ways with a shorter pathway for the in-plane exchange interaction, the nearest-neighbor in-plane exchange interaction $J_{ab}$ (previously quoted $J_1$) is expected to be much higher than the out-of-plane exchange interaction $J_c$ (*i.e.*, $J_{ab} \gg J_c$). Interestingly, the analysis of the susceptibility data also yields an exchange interaction along the *c* axis much weaker than that in the *ab* plane ($J_c / J_{ab} \sim 0.03$).[6] The stronger in-plane interaction leads to the building up of the spin-spin correlations in the *ab* plane at a temperature (~7 K) well above the 3D ordering temperature (3.75 K). A short-range order in the spin 1/2 AFM Heisenberg model with spatially anisotropic couplings was theoretically predicted for quasi 2D systems.[8]

Now we focus on another important result that we have observed in this present study. Below $T_N$, an asymmetric broad peak, centred around $Q \sim 0.7$ Å$^{-1}$, has been observed to coexist with the magnetic Bragg peaks down to 1.5 K (lowest measured temperature). The magnetic scattering profiles at two representative temperatures (1.5 and 2 K) are shown in Figs. 6(a) and 6(b). Here, a broad asymmetric peak is present along with the 3D magnetic Bragg peak (100). The peak position and peak shape are quite similar to those observed above $T_N$ due to a 2D magnetic ordering. However, the coexistence of 2D short-range and 3D long-range magnetic orderings is not expected in classical 2D systems like K$_2$CoF$_4$ ($T_N$ = 107.72 K)[17] and K$_2$NiF$_4$ ($T_N$ = 97.23 K),[18] where the 2D magnetic ordering fully converts to 3D ordering at $T_N$ so that the diffuse broad peak disappears when the sharp magnetic Bragg peaks appear at $T_N$. In the present study, we attribute the observed broad peak below $T_N$ to the contribution of spin-waves which are expected to be strongly enhanced for a spin 1/2 system close to 2D behaviour, such as Zn$_2$VO(PO$_4$)$_2$. The influence of spin-waves is also manifested in the reduced value of the ordered moment [0.66(2) $\mu_B$ per V ion] at 1.5 K, as discussed earlier.

To confirm this interpretation, we have calculated the spin wave spectrum for the quasi 2D spin 1/2 Heisenberg system Zn$_2$VO(PO$_4$)$_2$ , taking into account not only the planar near neighbour (NN) interaction $J_1$, but also the planar next nearest neighbour (NNN) interaction $J_2$, the interlayer interaction $J_\perp$, and the uniaxial anisotropy of the V$^{4+}$ ion. All these parameters have different impacts on the spin wave dispersion and ordered moment value, which we discuss first of all qualitatively. In a square lattice, spin frustration arises from the competition of $J_1$ and $J_2$ AFM exchange interactions [Fig. 3(b)], leading to a reduction of the ordered moment and to the possible disappearance of the long range magnetic order. This case is extensively studied in the $J_1$–$J_2$ model,[7] for frustrated spin-1/2 Heisenberg antiferromagnets on the stacked square-lattice, which shows several interesting magnetic ground states involving quantum phase transitions, such as, (i) for $\alpha$ (= $|J_2/J_1|$) $\leq$ 0.4, a Néel AFM state with nearest neighbor AFM coupling of spins in a given plane, as found for the present compound, (ii) for $\alpha \sim$ 0.4-0.6, a quantum spin-liquid state and, (iii) for $\alpha \geq$ 0.6, an ordered collinear AFM state i.e, AFM coupling of ferromagnetic chains in a given plane (stabilized an order by disorder process ). The interlayer interaction $J_\perp$ tends to restore a 3D behavior and therefore to increase the ordered moment. The sign of $J_\perp$ is important since a FM $J_\perp$ should reduce the zero point fluctuations more than an AFM $J_\perp$. Therefore, one expects the ordered moment to increase more for a FM $J_\perp$ than for an AFM $J_\perp$. The uniaxial anisotropy ($D$) creates a gap in the spin waves spectrum, and therefore induces a suppression of the low energy spin fluctuations. This in turn stabilizes the value of the ordered magnetic moment. To get a quantitative insight on the influence of all these factors on the spin wave spectrum, we have modelled the spin wave dispersion curves, by writing the energy in a way similar to that done in the compound LiFePO4 by Li *et al.*[19]

The Hamiltonian for the present spin system has been considered as

$$H = J_1 \sum_{i,\delta}\left(S_i S_{i+\delta}\right) + J_2 \sum_{i,\xi}\left(S_i S_{i+\xi}\right) + J_\perp \sum_{i,\delta_\perp}\left(S_i S_{i+\delta_\perp}\right) - D \sum_i \left(S^z\right)^2 \quad (2)$$

where, $J_1$ and $J_2$ are in-plane NN and NNN exchange interaction constants, respectively, $J_\perp$ is the inter-plane exchange interaction constant,, and $D$ is the uniaxial anisotropy constant which quantifies the tendency of the spins to align along the easy axis (the quantization axis is defined to be along the moment direction in the ground state, i.e., along the *c*-axis in the present system). Using the AFM spin wave theory,[20, 21] the lattice with N sites is divided into two sub lattices, namely, spin up (0, 0, $z$) and spin down (0.5, 0.5, z). The magnon dispersion curves were calculated using the Holstein-Primakoff spin operator transformation in the linear approximation i.e., linear spin-wave theory.[9, 22] The resulting spin-wave dispersion is given by

$$\hbar\omega = \sqrt{A^2 - F^2} \qquad (3)$$

Where

$A = 2J_1 ZS - 2J_2 ZS - 2J_\perp ZS + 2J_2 ZS\gamma_{NNN} + 2J_\perp ZS\gamma_\perp + 2DS$ and $F = 2J_1 ZS\gamma_{NN}$ in which Z is the number of the nearest neighbors (Z = 4 for intra-planner interactions and Z = 2 for inter planner interactions) and S is the spin value (S = 1/2 for $V^{4+}$). $\gamma_{NN}$, $\gamma_{NNN}$, and $\gamma_\perp$ are calculated using the following equation

$$\gamma(NN, NNN, \perp) = \frac{1}{Z}\sum_r e^{iQ \cdot r} \qquad (4)$$

Where, $r = (r_{NN}, r_{NNN}, r_\perp)$ are the lattice vectors between a given moment and its intra-plane nearest, next-nearest neighbor and inter-plane nearest-neighbor moments. For $Zn_2VO(PO_4)_2$ with tetragonal crystal structure (space group $I4cm$), the $V^{4+}$ ions are situated at 4a (0, 0, z) position. Therefore, from Eq. 4 we get

$$\gamma_{NN} = \cos(\pi k_x)\cos(\pi k_y) \qquad (5)$$

$$\gamma_{NNN} = \frac{1}{2}\left[\cos(2\pi k_x) + \cos(2\pi k_y)\right] \qquad (6)$$

and $\gamma_\perp = \cos(\pi k_z)$ $\qquad (7)$

The present calculation may help to explain, at least partly, the observed diffuse scattering at $T < T_N$. To this aim, we calculate the spin wave contribution to the "instantaneous" spin correlations function measured in diffraction experiments. In such experiments, the energy $E_f$ of the scattered neutrons spreads over the energy range from zero to infinity. However, at $T = 0$, neutrons can only create spin wave excitations, restricting this range up to $E_f = E_i$ (the incident energy of the neutrons). As a result, the inelastic contribution probes the spin wave spectrum from $\omega = 0$ to $\omega = E_i$. More precisely, for a given $Q$ modulus, the energy range is defined by the scattering laws. A reasonable agreement is obtained [see Figure 6(a) and 6(b)] between experimental spectra at 1.5 and 2 K, and spin wave calculation, by considering the values of $J_1$ = 7.9 K (AFM), $J_2$ = 0.2 K (AFM), $J_\perp$ = -0.2 K (FM), and $D = -0.1$, proposed by Kini *et al*. from their dc susceptibility study.[6]. The calculated spin wave spectra along the ($hh0$) and ($00l$) directions at $T = 0$ K are shown in Figs. 7(a) and 7(b), respectively, while Figs 7(c) and 7(d) show the inelastic and integrated powder averaged spectra. Note that the spin wave calculation is done in a semi-classical manner, by reference to an ordered state, and it assumes that the spin waves are only a perturbation of this state. The model is probably too rough for the present system $Zn_2VO(PO_4)_2$ and we plan to carry out an inelastic neutron scatting study to confirm the detailed nature of spin-waves in this compound.

A phase diagram for the studied quasi-2D compound $Zn_2VO(PO_4)_2$ is proposed in Fig. 8. With a lowering of temperature, the $\xi_{2D}$ increases monotonically down to $T_N$. The true 3D LR AFM ordering (along with spin-waves) is found only at $T \leq 3$ K ($T_N = 3.75$ K) where the widths of the magnetic Bragg peaks are limited by the instrumental resolution of the G6.1 diffractometer. Over the intermediate temperature range [3 K < $T$ < $T_N$ (3.75 K)], a broadening of the magnetic Bragg peaks is observed indicating a decrease of the 3D correlation length. This type of ordering could be termed as a 3D restricted long-range (RLR) AFM ordering.

## IV. SUMMARY AND CONCLUSION

In summary, the quasi 2D compound $Zn_2VO(PO_4)$ shows a rich magnetic phase diagram of pure 2D short-range ordering above $T_N$ (up to 7 K) and, 3D LR ordering along with spin-waves below $T_N$. The observed 3D long-range ordering is Néel type AFM ordering in the *ab* plane. A reduced value of the ordered moment [0.66(2) $\mu_B$ per V ion at 5 K] is found. The AFM layers are coupled ferromagnetically along the *c* direction. The observed Néel type long-range AFM ordering (with reduced moment) in the *ab* plane is in consistent with the theoretically predicted ground state for a frustrated spin-1/2 square-lattice AFM system with a weak next-nearest-neighbor exchange interaction and weak interlayer coupling. The observed in-plane (2D) short-range magnetic ordering above $T_N$ is in consistent with the theoretical predictions for quasi-2D systems.[8] This type of 2D magnetic ordering arises due to the anisotropic exchange pathways giving a stronger in-plane exchange coupling. The calculation of the magnetic scattering by spin-waves agrees rather well with the observed diffuse neutron scattering below $T_N$. The present study would be useful to understand magnetic properties of low dimensional systems, in general.

**ACKNOWLEDGMENT** The authors are grateful to C. Lacroix and the referee for fruitful discussions and valuable comments, respectively.

**FIGURE CAPTIONS**

FIG. 1. (Color online) Rietveld refined neutron diffraction patterns of $Zn_2VO(PO_4)_2$ at (a) 300 K and (b) 1.5 K. The observed and calculated diffraction patterns are indicated by circles and solid line, respectively. Solid line at the bottom shows the difference between observed and calculated patterns. Tick marks show the positions of nuclear Bragg peaks (space group: I4cm) and magnetic Bragg peaks (space group: P-1) (bottom row corresponding to the 1.5 K pattern only). The magnetic peaks are labelled with asterisks. The upper inset in the panel (b) shows an enlarged view of the diffraction pattern at 1.5 K. The lower inset in the panel (b) shows the temperature dependence of the ordered magnetic moment of V ion. (c) Magnetic pattern at 1.5 K, after subtraction of the pattern at 15 K (paramagnetic state), along with the patterns calculated by considering either a ferromagnetic (black line) or an antiferromagnetic (red squares) coupling along the $c$ direction. Tick marks show the Bragg peak positions considering a magnetic unit cell same as chemical unit cell.

FIG. 2. (Color online) (a) Crystal structure of the quasi-2D compound $Zn_2VO(PO_4)_2$. (b) Projection of the crystal structure in the $ab$ plane. The dimmers of $ZnO_5$ square pyramides are evident.

FIG. 3. (Color online) (a) Magnetic structure of $Zn_2VO(PO_4)_2$. The origin is shifted along the positive $c$- axis by an amount of $z/c = 0.1174$. (b) The Néel type AFM ordering of the spins in a given $ab$ plane. The nearest-neighbour and next-nearest-neighbour exchange interactions are labelled $J_1$ and $J_2$, respectively.

FIG. 4. (Color online) (a)-(d): Magnetic diffraction patterns at 10, 7, 4.5, and 4 K, respectively, after subtracting the nuclear background at 15 K from all patterns. The solid curves [(a)-(d)] are the profiles calculated using the Warren function expected for a purely 2D magnetic ordering.

FIG. 5. (Color online) The in-plane (V-O1-P-O1-V) and out-of-plane (V-O1-Zn-O2-P-O1-V) shortest possible magnetic exchange interaction pathways.

FIG. 6. (Color online) Magnetic diffraction patterns at (a) 1.5 K and (b) 2 K, after subtracting out the nuclear background at 15 K. The thin curves are the 3D Bragg peaks calculated by considering a Pseudo-Voigt function for the resolution profile. The open circles are the calculated profile of the spin-waves contribution to the scattering (with the values of $J_1$, $J_2$, $J_\perp$ and D of 7.9 K (AFM), $J_2 = 0.2$ K (AFM), $J_\perp = -0.2$ K (FM), and $D = -0.1$, respectively). The thick red curves are the sum of the spin-waves and Pseudo-Voigt (3D) peak profiles.

FIG. 7. (Color online) Calculated spin-wave spectrum along (a) $(h, h, 0)$ and (b) $(0, 0, l)$ directions. (c) The powder averaged spin-wave spectrum. (d) The powder averaged $S(Q)$ vs. $Q$ (with the values of $J_1$, $J_2$, $J_\perp$ and D of 7.9 K (AFM), $J_2 = 0.2$ K (AFM), $J_\perp = -0.2$ K (FM), and $D = -0.1$, respectively).

FIG. 8. (Color online) Temperature dependence of FWHM for the 3D magnetic Bragg peak (100) and the 2D spin-spin correlation length ($\xi_{2D}$). Several magnetic ordering regions, 3D long-range with spin-waves, 3D reduced long-range with 2D short-range, and 2D short-range, are indicated.

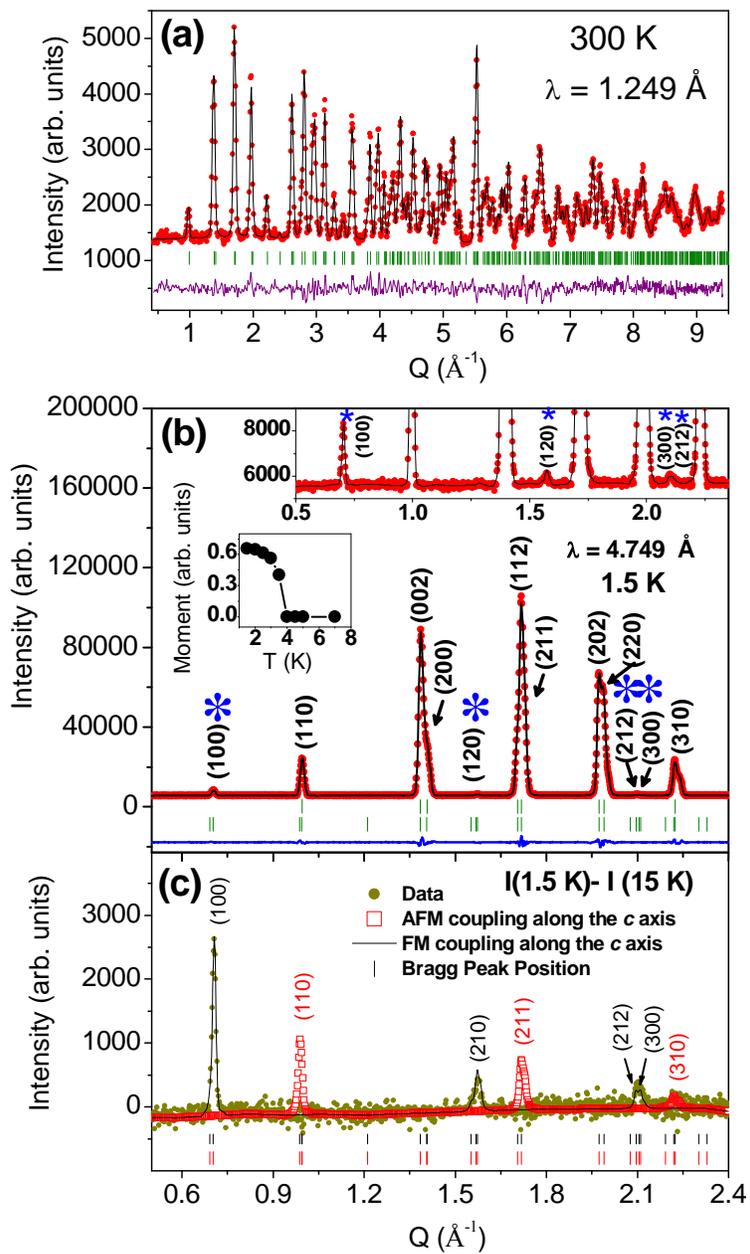

Fig. 1    Yusuf *et al.*

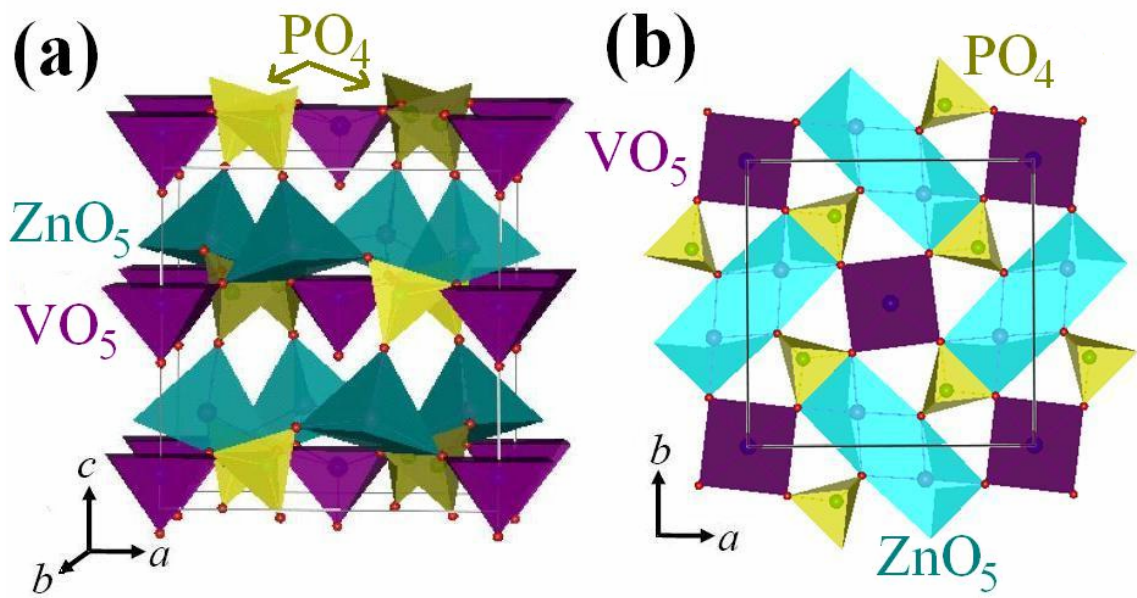

Fig. 2          Yusuf *et al.*

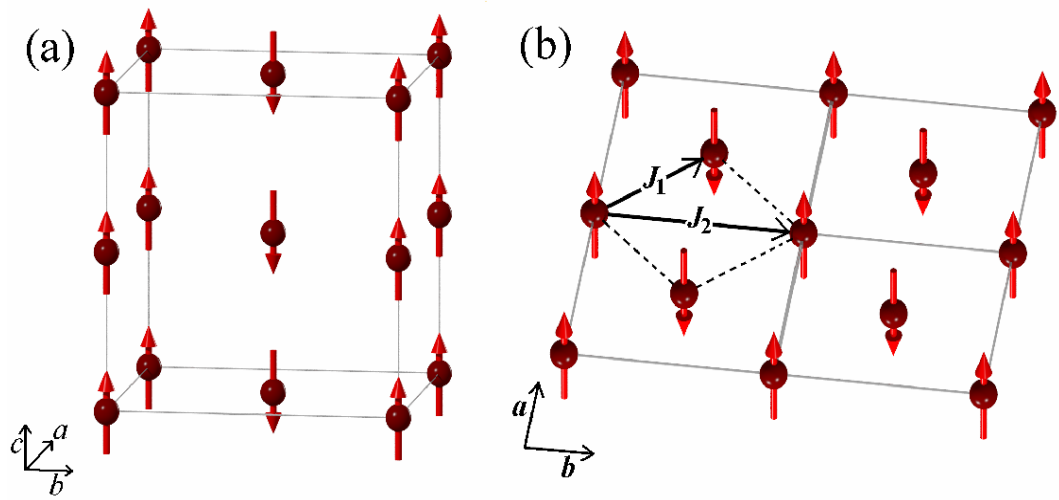

Fig. 3     Yusuf *et al*.

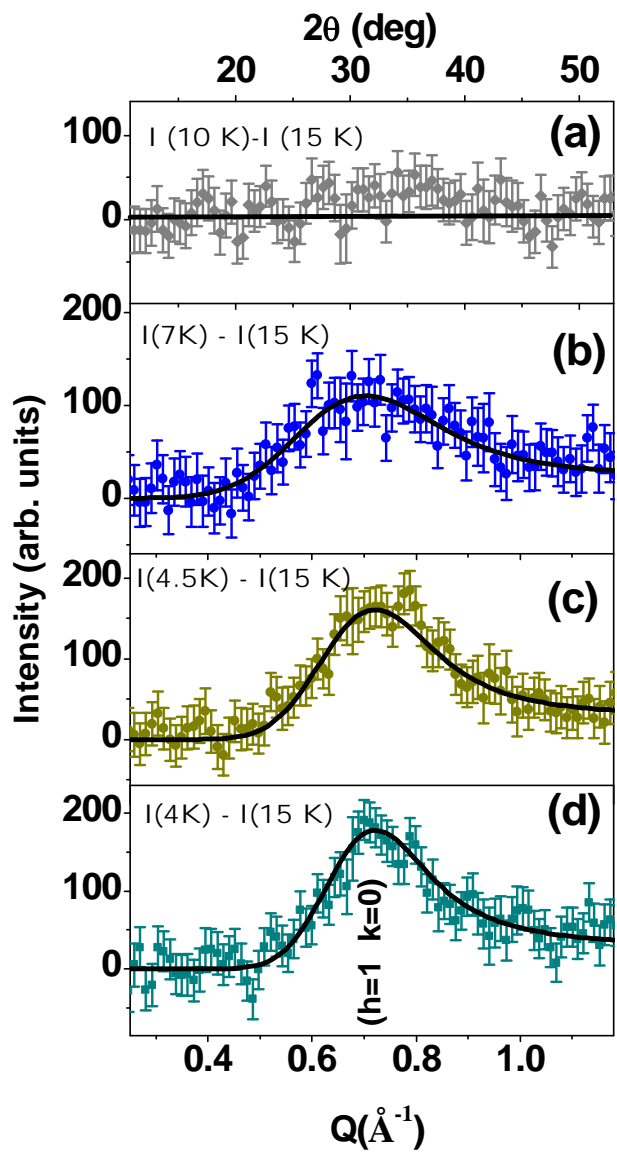

Fig. 4    Yusuf *et al.*

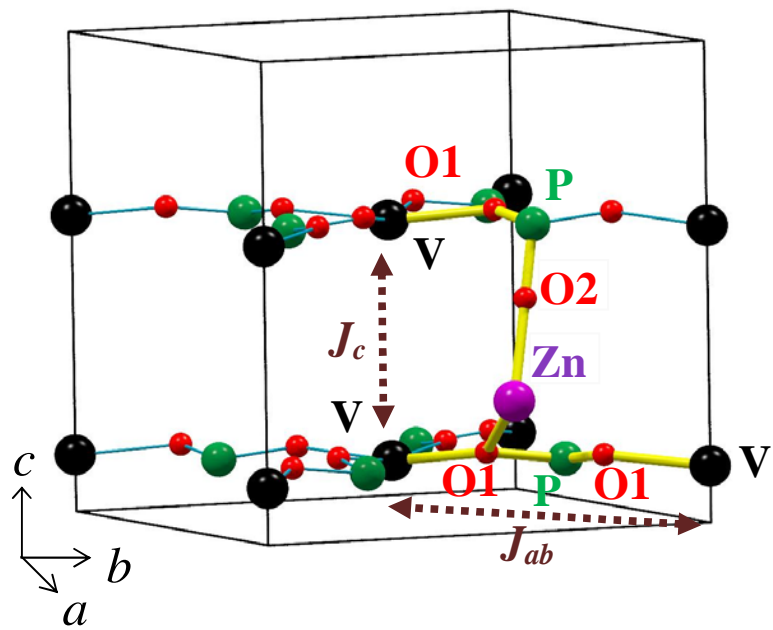

Fig. 5          Yusuf *et al.*

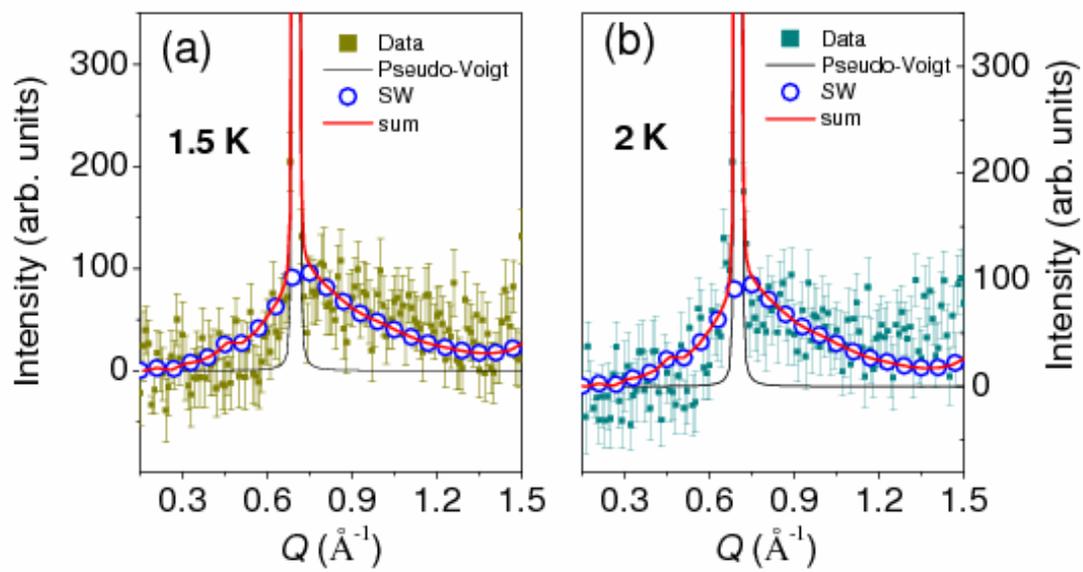

Fig. 6          Yusuf *et al.*

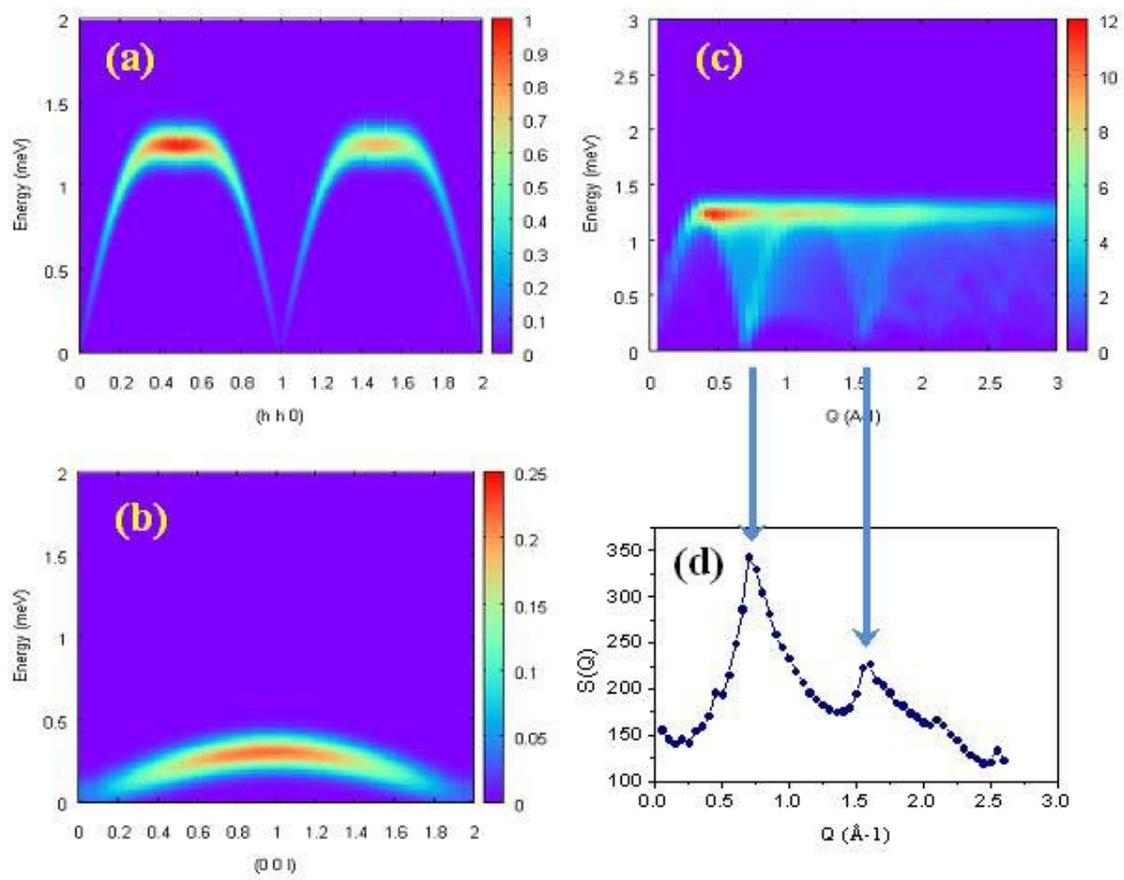

Fig. 7    Yusuf *et al.*

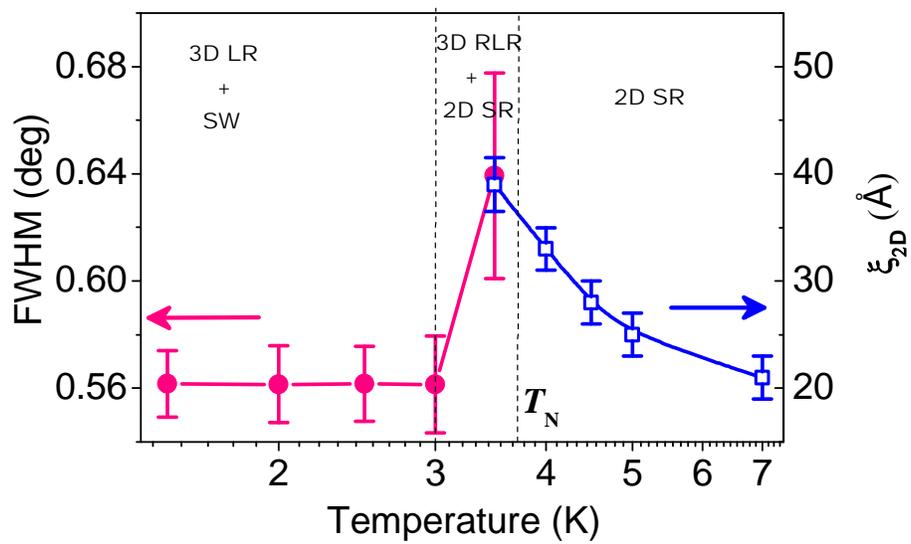

Fig. 8            Yusuf *et al.*